\documentclass[referee]{raa}

\usepackage{graphicx,times}
\usepackage{natbib}
\usepackage{amssymb,amsmath}
\bibpunct{(}{)}{;}{a}{}{,}
\usepackage{rotating}
\usepackage{pdflscape}

\usepackage[pagebackref=true]{hyperref}

\begin{document}

   \title{The relative orientation between local magnetic field and Galactic plane in low latitude dark clouds}

 \volnopage{ {\bf 20XX} Vol.\ {\bf X} No. {\bf XX}, 000--000}
   \setcounter{page}{1}

   \author{Gulafsha B. Choudhury \inst{1}, Himadri S. Das\inst{1}, B. J. Medhi \inst{2}, J. C. Pandey \inst{3}, S. Wolf
      \inst{4}, T. K. Dhar\inst{1}, A. M. Mazarbhuiya\inst{1}
   }

   \institute{Department of Physics, Assam University, Silchar 788011, India; {\it himadri.sekhar.das@aus.ac.in} and {\it gulafsha.97@gmail.com}\\
        \and
             Department of Physics, Gauhati University, Guwahati 781014, India\\
	\and
Aryabhatta Research Institute of Observational Sciences (ARIES), Nainital 263002, India\\
\and
University of Kiel, Institute of Theoretical Physics and Astrophysics, Leibnizstrasse 15, 24118, Kiel, Germany\\
\vs \no
   {\small Received 20XX Month Day; accepted 20XX Month Day}
}

\abstract{In this work, we study the magnetic field morphology of selected star-forming clouds spread over the galactic latitude ($b$) range, $-10^\circ$ to $10^\circ$.
The polarimetric observation of clouds CB24, CB27 and CB188 are conducted to study the magnetic field geometry of those clouds using the 104-cm Sampurnanand Telescope (ST) located at ARIES, Manora Peak, Nainital, India.
These observations are combined with those of 14 further low latitude clouds available in the literature. Most of these clouds are located within a distance range 140 to 500 pc except for CB3 ($\sim$2500 pc), CB34 ($\sim$1500 pc), CB39 ($\sim$1500 pc) and CB60 ($\sim$1500 pc). Analyzing the polarimetric data of 17 clouds, we find that the alignment between the envelope magnetic field ($\theta_{B}^{env}$) and Galactic plane ($\theta_{GP}$) of the low-latitude clouds varies with their galactic longitudes ($l$). We observe a strong correlation between the longitude (\textit{l}) and the offset ($\theta_{off}=|\theta_B^{env}-\theta_{GP}|$) which shows that $\theta_{B}^{env}$ is parallel to the Galactic plane (GP) when the clouds are situated in the region, $115^\circ<l<250^\circ$. However, $\theta_{B}^{env}$ has its own local deflection irrespective of the orientation of $\theta_{GP}$ when the clouds are at $l<100^\circ$ and $l>250^\circ$. To check the consistency of our results, the stellar polarization data available at Heiles (2000) catalogue are overlaid on DSS image of the clouds having mean polarization vector of field stars. The results are almost consistent with the Heiles data. A systematic discussion is presented in the paper. The effect of turbulence of the cloud is also studied which may play an important role in causing the misalignment phenomenon observed between $\theta_{B}^{env}$ and $\theta_{GP}$. We have used \textit{Herschel}\footnote{Herschel is an ESA space observatory with science instruments provided by European-led Principal Investigator consortia and with important participation from NASA.} \textit{SPIRE} 500 $\mu m$ and \textit{SCUBA} 850 $\mu m$ dust continuum emission maps in our work to understand the density structure of the clouds.
}

   \authorrunning{G. B. Choudhury et al.}            
   \titlerunning{Magnetic field morphology of clouds}  
   \maketitle

%
\section{Introduction}

Magnetic fields are present everywhere in our galaxy, spreading the interstellar medium and expanding beyond the Galactic disk. They are present in a broad variety of astrophysical objects, such as, molecular clouds, pulsars, supernova remnants \citep{Lu2020}. Various astronomers extensively studied the large-scale galactic magnetic field (GMF), yet it remains inadequately understood. The GMF plays an essential role in forming molecular clouds that serve as the stellar nest in our galaxy. Galactic fields could be sufficiently strong to inflict their direction upon individual molecular clouds \citep{shetty2006global}, that can modulate the accumulation and fragmentation of the cloud \citep{li2011evidence}, thereby altering the efficiency of star formation \citep{price2008effect}. The magnetic field in molecular clouds plays a significant role in star formation efficiency \citep{Hennebelle2019}. Magnetic fields are also believed to have a considerable impact on the circumstellar disc formation as well as on fragmentation in forming binary systems \citep{price2007impact}. There are various other parameters responsible for star formation processes that involves turbulence \citep{li2004formation, tilley2004formation, vazquez2005star}, jets and feedback from outflows \citep{li2006cluster, Vazquez2019, 2019FrASS...6...54P}, radiation feedback from the stars themselves \citep{clark2005star}. There is evidence of molecular clouds showing turbulent motions \citep{larson_turbulence_1981, mac2004control}. The impact of turbulence on the magnetic field structure is generally tough to interpret. However, some studies show that the magnetic field may play a dominant role in shaping the dynamics of the turbulence \citep{padoan2002stellar, padoan2007two, padoan2014star}. Thus, it is important to study the magnetic field morphology to understand the ongoing activities in molecular clouds.

When the background starlight passes through the aligned dust grains present in the interstellar medium, it gets polarized and the polarization position angle gives the orientation of local magnetic field. \citet{draine_radiative_1996} suggested that such alignment of the dust grains present in the molecular clouds may be due to the effect of radiative torque. The radiative torque mechanism is established on the interaction between radiation and grain to spin it up. The confirmations on radiative torque alignment (RAT) was established by \citet{whittet_interstellar_2001} while studying the dense and diffuse gas at the Taurus cloud. In recent years, diverse studies were made on grain alignment by RAT mechanism, which reveals that RAT happens to be a successful mechanism alignment that can explain the dust grain alignment of numerous astrophysical environments \citep{hoang_grain_2014, hoang_modelling_2015, andersson_interstellar_2015}. \citet{hoang_grain_2014} found that the linear polarization of nearby stars as predicted by the radiative alignment torque agrees well with the observational data, which manifests that polarization increases with the distance to the stars. \citet{andersson_interstellar_2015} mentioned that the theory of interstellar grain alignment by RAT allows deriving specific, testable predictions for practical interstellar processes. Further detailed analysis of the RAT mechanism might give a promising explanation of grain alignment and polarimetry on the interstellar magnetic field and provide advanced information on dust characteristics.

Several researchers studied the orientation of magnetic field through imaging polarimetry  \citep{chakraborty_study_2014, soam_magnetic_2015, chakraborty_study_2016, das_magnetic_2016, soam_probing_2017, jorquera_magnetic_2018, choudhury_bok_2019, zielinski_constraining_2021} and discussed the relative orientation of magnetic field to the Galactic plane (GP), outflow direction and minor axis of the cloud. The optical polarimetric analysis reveals that the envelope magnetic field of CB130 is oriented at an angle of 53$^\circ$ with respect to the orientation of GP \citep{chakraborty_study_2016}. \citet{das_magnetic_2016} estimated the magnetic field strength of two sub-mm cores of CB34 from the archival sub-millimeter polarimetric data. They presented the relative orientation of the envelope magnetic field with the minor axis of the cloud for both the cores, which supports magnetically dominated star formation models. Analyzing both optical and sub-mm polarimetric data of Bok globule CB17, \citet{choudhury_bok_2019} reported a parallel alignment between envelope magnetic field and the position angle of GP in contrast to the core-scale magnetic field, which is almost perpendicular to the GP. They also reported a relative orientation envelope magnetic field to the outflow axis and the cloud minor axis. \citet{zielinski_constraining_2021} discussed the magnetic field of a prototypical cloud B 335 and observed a decrease in polarization towards the center of the cloud (dense core). They also observed a uniform pattern in the polarization vectors.

In this article, we present the magnetic field morphology of 17 star forming clouds (including newly observed clouds CB24, CB27 and CB188), spread over the low galactic latitude range of $-10^\circ < b < 10^\circ$. Most of these clouds are located within a distance range 140 to 500 pc except for CB3 (2500 pc), CB34 (1500 pc), CB39 (1500 pc) and CB60 (1500 pc). We systematically study the alignment mechanism between the envelope magnetic field of the cloud and Galactic plane (GP) and their variation with galactic longitude. In \S \ref{s2}, we present the observation, data reduction procedures along with the details of the archival data. In \S \ref{s3}, we discuss the geometry of the envelope magnetic field of the three observed clouds. We summarize our results in  in \S \ref{s4}.


\begin{table*}
\caption{Observation log.}
\begin{center}
\begin{tabular}{cccccc}
\hline
\hline
   Object ID & Name of         &Date           & Fields & RA(2000) & DEC(2000)           \\
             & Observatory     &               &        & (h m s)  & ($^\circ$ $^\prime$ $^{\prime\prime}$)  \\
 \hline
 \hline
   CB24     & ARIES, Nainital &  Dec 23, 2017 &  F1    & 04:58:30  & 52:12:17 \\
             &                 &               &  F2    & 04:58:23  & 52:16:31 \\
			 &                 &               &  F3    & 04:59:03	& 52:17:02 \\
			 &                 &               &  F4    & 04:59:00  & 52:10:49 \\
 \hline
   CB27     & ARIES, Nainital &  Dec 22, 2017 &  F1    & 05:04:07  & 32:38:19 \\
             &                 &               &  F2    & 05:03:31	& 32:42:16 \\
			 &                 &               &  F3    & 05:03:33  & 32:48:51 \\
			 &                 &               &  F4    & 05:04:09  & 32:51:33 \\
\hline
   CB188    & ARIES, Nainital &  May 8, 2019  &  F1    & 19:20:25  & 11:33:10 \\
             &                 &               &  F2    & 19:20:29	& 11:40:33 \\
			 &                 &               &  F3    & 19:19:59	& 11:32:06 \\
			 &                 &               &  F4    & 19:20:06  & 11:39:56 \\
 \hline
\end{tabular}
\end{center}
\label{t}
\end{table*}

\section{Description of sources}
\label{s2.2.2}
\subsection{CB24}

CB24 is a starless, small spherical cloud at a distance of $293 \pm 54$ pc \citep{das_distance_2015}. There was no association with an IRAS point source is identified. \citet{kane_search_1994} found that CB24 is a considerably less dense cloud, and the low column density may indicate that Bok globules as CB24 did not undergo significant core contraction and represent an ideal sample of starless small dark clouds.

\subsection{CB27}

CB27, also known as L1512, is an isolated Taurus core cloud near the GP. The distance of  CB27 is found to be 140 pc \citep{kenyon_new_1994}. A compact submillimeter source (FWHM $\sim$ 104 AU) is found to be present at the center of CB27 \citep{di_francesco_scuba_2008, kirk_initial_2005}. The central density of the cloud is near to the maximum stable density, which is required for a pressure-supported, self gravitating cloud, and this makes the cloud indistinguishable whether the core is a starless stable or a prestellar one \citep{launhardt_earliest_2013}.

\subsection{CB188}
\label{s2.3}
CB188 is an isolated small cloud at a distance $262 \pm 49$ pc \citep{das_distance_2015}. The bolometric luminosity of this cloud is 2.6 $L_{\odot}$ and the envelope mass is of about 0.7 $M_{\odot}$ obtained from an interferometric study of the $N_2H^+$ (1--0) emission \citep{chen_ovro_2007}. The mean density of the core is $\sim 2 \times 10^6$ $cm^{-3}$. CB188 is found to be physically associated with L 673 \citep{tsitali_spitzer_2010}, as shown by the dotted rectangle in the lower region of Figure~\ref{fig3}. In our study, we only covered the northern region of the cloud CB188.

\section{OBSERVATION, DATA REDUCTION and ARCHIVAL DATA}
\label{s2}
\subsection{Observations}
\label{s2.1}
 We conducted the optical polarimetric observations of four fields towards the Bok Globules CB27, CB24 and CB188 each on  December 22, 2017, December 23, 2017 and May 8, 2019, respectively. We have selected these three globules because of their close proximity with the GP ($-10^\circ < b < 10^\circ$) as we aim to study the magnetic field morphology of low latitude clouds. Also, no polarimetric study of these clouds was performed in the past. Moreover, the availability of dust continuum emission maps of CB27 (\textit{Herschel SPIRE} 500 $\mu m$) and CB188 (and \textit{SCUBA} 850 $\mu m$) also motivated us to select these globules. Polarimetric observations were conducted in the R-band (filter: $\lambda$ = 630 nm, $\Delta \lambda$ = 120 nm). The observed region around each globule is divided into four fields of 8$^\prime\times$8$^\prime$ dimension because the CCD has a diameter field of view of $\sim$ 8$^\prime$. The observations were carried out with the 104 cm Sampurnanand Telescope (ST) at Aryabhatta Research Institute of observational sciencES, Nainital, India  \citep[for exemplary previous polarimetric observations using this telescope see][]{chakraborty_study_2014, das_magnetic_2016, chakraborty_study_2016}. The observation log is presented in Table~\ref{t}. The 104 cm ST is an f/13 Cassegrain telescope. An Imaging Polarimeter (AIMPOL) is connected to the back-end of the telescope that has a Wollaston prism and a rotating half-wave plate (HWP). The Wollaston prism splits the incoming unpolarized light into two orthogonal components (ordinary and extraordinary), and the HWP rotates the polarization state of light into four angles $0^\circ$, $22.5^\circ$, $45^\circ$ and $67.5^\circ$ which gives the four polarized components \citep[see][for detailed observational procedures]{das2013polarimetric}.  The detailed theory and design of AIMPOL are presented in \citet{Rautela2004} and \citet{medhi2008optical}.


\begin{table*}
\caption{Standard star polarimetry: Object ID, date of observation, observed values of $p$ and $\theta$, literature values of $p$ and $\theta$, reference.}
\begin{center}
\scriptsize{
\begin{tabular}{llcccccc}
\hline
\hline
&& \multicolumn{2}{c}{Observed values}  && \multicolumn{2}{c}{Literature values} &\\
\cline{3-4} \cline{6-7}
&&&&&&\\
Object ID & Date of observation & $p$ $\pm$ $e_p$  & $\theta$ $\pm$ $e_{\theta}$ & & $p$ $\pm$ $e_p$  & $\theta \pm e_{\theta}$ & Literature reference \\
   &  & ($\%$) & ($^\circ$) && ($\%$) & ($^\circ$) &   \\
 \hline
 \hline
 & &\multicolumn{4}{c}{\textit{High polarized standard stars}} & \\
\hline

HD 251204   & December 23, 2017 & 4.84$\pm$0.17     &   154.3$\pm$1.0   &&   4.79$\pm$0.30  & 155.7  & \citet{serkowski1974polarimeters}  \\
HD 19820    & December 23, 2017 & 4.83$\pm$0.18     &   115.4$\pm$1.0   &&   4.53$\pm$0.03  & 114.5$\pm$0.2 & \citet{schmidt1992hubble}  \\
HD 154445   & May 8, 2019       &  3.48$\pm$0.09    &   89.7$\pm$0.7    & &  3.78$\pm$0.06  & 88.8$\pm$0.5  & \citet{schmidt1992hubble} \\
HD 161056   & May 8, 2019       &  3.85$\pm$0.10    &   69.5$\pm$0.7    &  & 4.03$\pm$0.03  & 66.9$\pm$0.2 & \citet{schmidt1992hubble} \\
\hline
\hline
& &\multicolumn{4}{c}{\textit{Unpolarized standard stars}} & \\
\hline
HD 21447    & December 23, 2017 & 0.11$\pm$0.18     &   111.7           & &  0.06$\pm$0.03  & 110 & \citet{breeveld_search_1998} \\
$\gamma$Boo & May 8, 2019       &  0.19$\pm$0.12    &   23.1            & &  0.065$\pm$0.02 & 21.3 & \citet{schmidt1992hubble} \\
$\beta$UMa  & May 8, 2019       &  0.10$\pm$0.14    &   109.2           & &  0.009$\pm$0.02 & 107.8 & \citet{schmidt1992hubble} \\
 \hline
\end{tabular}
}
\end{center}
 \label{t1}
\end{table*}


\begin{table*}
\caption{Details of target globules (our observation along with archival references of polarimetric studies). Cloud ID, Right ascension (RA), Declination (DEC), Galactic longitude ($l$), Galactic latitude ($b$), Name of observatory, Date of observation, Reference and distance to the cloud (d).}
\begin{center}
\tiny{
\begin{tabular}{cccrrcccc}
\hline
\hline
 ID & RA(2000) & DEC(2000)  & $l$ & $b$ &  Name of  &  Date of & Reference & Distance (ref) \\
		
	& (h m s) & ($^\circ$ $^\prime$ $^{\prime\prime}$)  & ($^\circ$)    & ($^\circ$) & observatory  & observation &  & (pc)  \\
 \hline
CB24	&	04 58 30	&	$+$52 15 41	& 155.76 &	5.90		&  $^{*}$ST	& Dec 23, 2017 & & 293$\pm$54 (1) \\
CB27	&	05 04 09	&	$+$32 43 12	& 171.82 &	$-$5.18	&  ST	& Dec 22, 2017 & Our observations & 140 (2) \\
CB188   &   19 20 17  &   $+$11 36 12    & 46.53  &   $-$1.01    &  ST & May 8, 2019 && 262$\pm$49 (1)  \\
\hline
&&& & & &  &\\
&&& & & & \textit{Archival Data}   &\\
&&& & & &  &\\
\hline
CB3		&	00 28 45	&	$+$56 42 08	& 119.80  &	$-$6.03	& $^{**}$MA & Dec 23, 1997 & \citet{sen_imaging_2000} & 2500 (3)	\\
CB4     &   00 39 03    &   $+$52 51 29    & 121.03 &   $-$9.96 &  $^{\dag}$MB   & Dec 1986  & \citet{kane_magnetic_1995} & 350$\pm$150 (4) \\
CB17	&	04 04 37	&	$+$56 56 41	& 147.02 &	3.39&	ST & March 9, 2016 & \citet{choudhury_bok_2019} & 253$\pm$43 (5) \\
CB25	&	04 59 04	&	$+$52 03 24	& 155.97 &	5.84	&  MA  & Dec 23, 1997  & \citet{sen_imaging_2000} & --  \\
CB26	&	05 00 09	&	$+$52 05 00	& 156.05 &	5.99	&	ST & Dec 29 \& 30,  & \citet{HalderCB26} & 140$\pm$20 (6)\\
&&&&&&2016&&\\
CB34	&	05 47 02	&	$+$21 00 10	& 186.94 &	$-$3.83	&	ST &  March 12–13, & \citet{das_magnetic_2016} & 1500 (3) \\
&&&&&&2013&&\\
        &               &               &       &           &                           &   Feb 20, 2015  &&   \\
CB39	&	06 01 58	&	$+$16 30 26	& 192.63 &	$-$3.04	&	MA & Dec 25, 1997 & \citet{sen_imaging_2000} & 1500 (3) \\
CB56	&	07 14 36	&	$-$25 08 54	& 237.90  &	$-$6.45	&	IGO & March 4, 2011 & \citet{chakraborty_study_2014} & -- \\
CB60	&	08 04 36	&	$-$31 30 47	& 248.89 &	$-$0.01	&	$^{\ddag}$IGO &  March 5, 2011 &  \citet{chakraborty_study_2014} & 1500 (3) \\

CB69 & 17 02 42 & $-33$ 17 00 & 351.23 & 5.14 & IGO & March 5, 2011 & \citet{chakraborty_study_2014} & 500 (3)\\

CB130	&	18 16 16	&	$-$02 33 01	& 26.61  &	6.65	&	IGO & April 26, 28   & \citet{chakraborty_study_2016} & 250$\pm$50 (3) \\
&&&&&& \& 30, 2014&&\\
        &               &               &       &           &                           &   May 2-4, 2014  & &  \\
CB246	&	23 56 44	&	$+$58 34 29	& 115.84 &	$-$3.54	&	MA & Dec 24, 1997  & \citet{sen_imaging_2000} & 140 (3)\\
L1014   &   21 24 07  &  $+$49 59 05    & 92.45  & $-$0.12  &   ST & Nov 14, 2010 & \citet{soam_magnetic_2015} & 258$\pm$50 (7) \\
        &               &               &       &           &     & Nov 22, 2011  &  & \\
L1415   &   04 42 00    &   $+$54 26 00    & 152.41 & 5.27     &  ST &  Nov 23, 2011  & \citet{soam_probing_2017} & 250 (8) \\
        &               &               &       &           &     & Dec 19, 20  &  & \\
        &&&&&& \& 24, 2011&&\\
        &               &               &       &           &     &  Oct 29, 2013 &   &\\

\hline
\end{tabular}
}
\end{center}
$^{*}$ST: Sampurnanand Telescope, ARIES, Nainital

$^{**}$MA: Mount Abu, India

$^{\dag}$MB: Mount Bigelow, North of Tucson, Arizona

$^{\ddag}$IGO: IUCAA Girawali Observatory, Pune

\textit{Distance References:} (1) \citet{das_distance_2015}. (2) \citet{kenyon_new_1994}. (3) \citet{launhardtHenning1997millimetre}. (4) \citet{perrot20033d}. (5) \citet{choudhury_bok_2019}. (6) \citet{launhardt_looking_2010}. (7) \citet{soam_magnetic_2015}. (8) \citet{soam_probing_2017}
 \label{t2}
\end{table*}


\subsection{Data reduction: Imaging polarimetry}
\label{s2.2}
 We conducted the observation using the four rotations of HWP as mentioned in \S \ref{s2.1}. For a particular rotation of HWP ($\alpha$), the intensities (extraordinary, $I_e$ and ordinary, $I_o$) of the two orthogonal polarized components are determined. If the HWP is rotated by $\alpha$, the electric vector rotates by 2$\alpha$. For calculation of the linear polarization it is useful to define the ratio $R_\alpha$
\begin{equation}
R_\alpha = \frac{(I_e/I_o) -1}{(I_e/I_o) +1} = p~ cos(2\theta-4\alpha)
\end{equation}
 \noindent where, $\theta$ and $p$ are the position angle and degree of linear polarization, respectively \citep{Rautela2004}. This ratio becomes $Q/I$ and $U/I$ when $\alpha= 0^\circ$ and $22.5^\circ$ respectively, i.e. the values of normalised Stokes parameters $q$ and $u$ ($I$: total intensity). The linear polarisation ($p$) and the polarization position angle ($\theta$) are given by
\begin{equation}
\label{eq1}
p = \sqrt{q^2 + u^2} \hspace{1cm} \text{and}\hspace{1cm} \theta = \frac{1}{2}tan^{-1}\bigg(\frac{u}{q}\bigg)
\end{equation}

In principle, the linear polarization and the position angle of polarization can be measured from the first two rotations of HWP. However, the additional two rotations $45^\circ$ and $67.5^\circ$ are observed due to non-responsivity of the system.

The observed polarimetric data have been reduced using the IRAF (Image Reduction and Analysis facility) package, \cite[see][for detailed data reduction procedures]{Rautela2004}. The uncertainties associated with $p$ and $\theta$ are calculated using the relations \citep{ramaprakash_imaging_1998}
\begin{equation}
e_p = \frac{\sqrt{N+N_b}}{N} ~~~~~ and~~~~~
e_\theta = 28.65^\circ \times \frac{e_p}{p}
\end{equation}

\noindent where $N$ and $N_b$ represent the flux counts corresponding to the source and background, respectively.


\begin{figure}[htb]
\begin{center}
	\vspace{-0.5cm}

	\includegraphics[width=\columnwidth]{{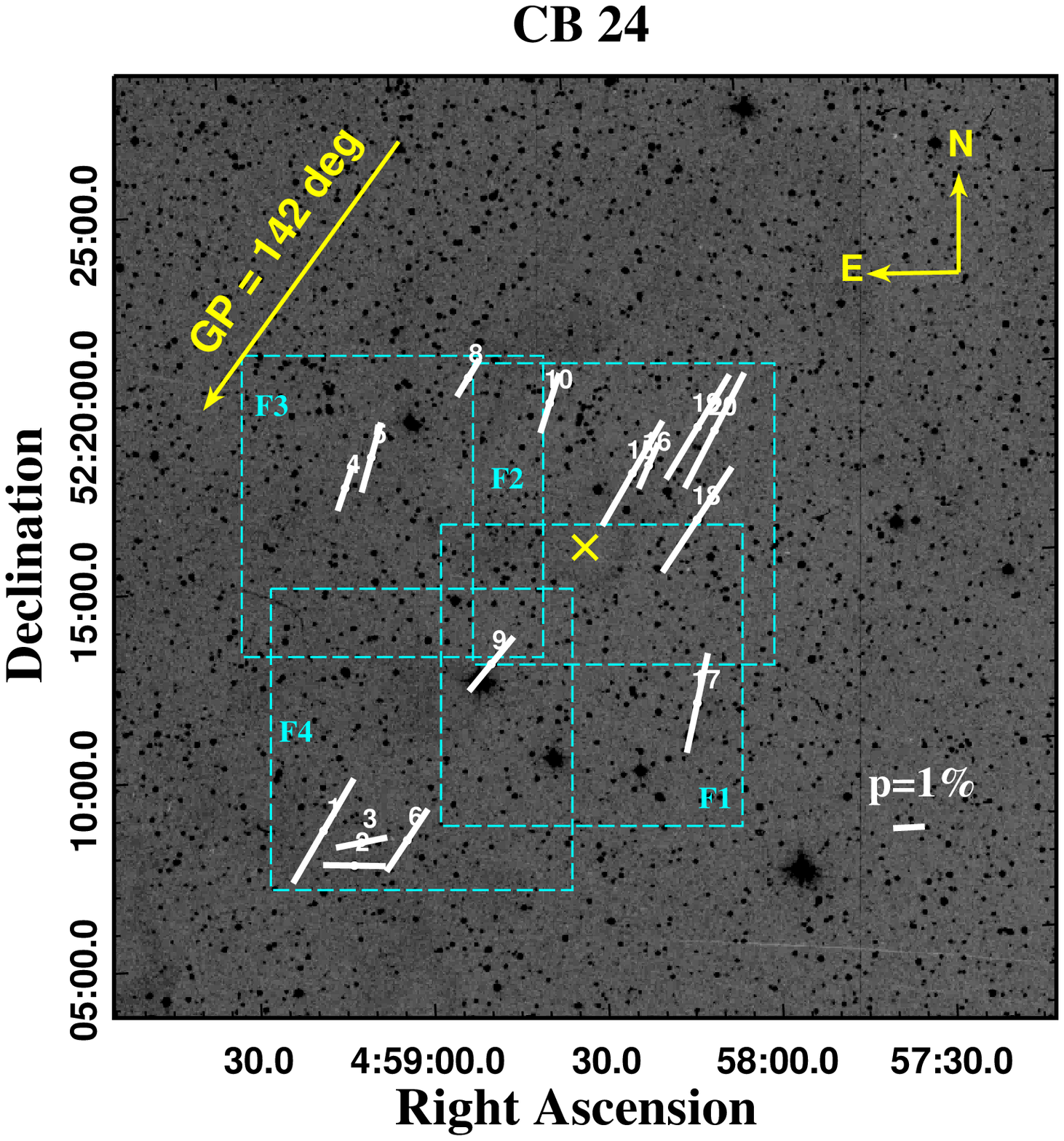}}
	\vspace{-1cm}
    \caption{Polarization map of CB24: White solid lines represent the polarization vectors of the background field stars plotted on a DSS image of the globule CB24 (25$^\prime\times25^\prime$). At the bottom right corner, a vector of 1$\%$ polarization is shown for reference. The vector at the top left corner shows the orientation of the GP ($\theta_{GP}=142^\circ$). The center of the globule is shown by the cross mark. The dashed rectangular boxes of dimension 8$^\prime\times8^\prime$ show the fields of observation (details are given in Table \ref{t}) of the cloud.}
    \label{fig1}
  \end{center}
\end{figure}


\subsubsection{Instrumental Calibration}
 The instrumental calibration is determined by analyzing three low polarized standard stars HD 21447, $\gamma$Boo, and $\beta$UMa taken from \citet{breeveld_search_1998} and \citet{schmidt1992hubble} which are in sound agreement with the literature. The instrumental calibration for zero position angle of polarization is determined by analyzing four high polarized standard stars HD 251204, HD 19820, HD 154445, and HD 161056 taken from \citet{serkowski1974polarimeters} and \citet{schmidt1992hubble}. The results obtained from our observations are presented in Table~\ref{t1}.

\subsection{Archival Data}
 Our observed clouds are situated at low galactic latitude close to GP, which allows us to map the magnetic field morphology of the star forming clouds near the GP. We collected additional 14 low galactic latitude star forming clouds for which polarimetric observations at optical wavelength are available in literature allowing us to perform a systematic statistical significant analysis.
These include 12 Bok globules (\textit{viz.} CB3, CB4, CB17, CB25, CB26, CB34, CB39, CB56, CB60, CB69, CB130 and CB246) and two Lynd's clouds (\textit{viz.} L1014 and L1415). All these clouds are located in the galactic latitude ($b$) ranging from $-10^\circ<b< 10^\circ$. Optical polarimetric observation of CB26 was performed by our group \cite[][in prep.]{HalderCB26}. The details of the 17 clouds are compiled in Table~\ref{t2}.

\section{Geometry of envelope magnetic field}
\label{s3}
 We reduce the optical polarimetric data of CB24, CB27 and CB188 using IRAF (as discussed in \S \ref{s2.2}). The values of $\theta$ and $p$ of the background stars detected towards the field of the three clouds are calculated using equation \ref{eq1}. We consider only those sources with $p/e_p\geq$ 3 (here $e_p$ denotes the polarization error). To avoid the foreground polarization, we make use of Gaia EDR3 parallaxes \citep{gaia_2016,gaia_2021} to determine the distance of the individual field stars towards each cloud. The critical distance is set to be the distance of the respective cloud ($293 \pm 54$ pc for CB24, 140 pc for CB27 and $262 \pm 49$ pc for CB188, respectively). For further analysis, we consider only sources with distances beyond the critical distance. In the case of CB24, polarization measurements of 20 fields stars are found, out of which three sources (\#7, \#12 and \#14) have been identified to be foreground to the cloud. Moreover, Gaia parallaxes for two sources (\#11 and \#13) are not available, so we discarded these five sources from further analysis. In the case of CB27, polarization measurements of 27 field stars are found, 26 of which have been identified to be background to the cloud while one source (\#16) does not have Gaia parallax available and hence we discard this star from the analysis as well. However, in the case of CB188, polarization measurements of 24 field stars are found and all these 24 sources have been identified to be background to the cloud.

Fifteen field stars are detected towards CB24, twenty six field stars towards CB27 and twenty four field stars are detected in the field of CB188, respectively. The values of $p$ and $\theta$ with the uncertainties of the field stars towards these three clouds are presented in Table~\ref{t3}, Table~\ref{t4}, and Table~\ref{t5}, respectively. The mean value of degree of polarization  ($<p>$) along with the standard error\footnote{$S.E = \frac{\sigma}{\sqrt{n}}$,
where $\sigma$ is the sample standard deviation and $n$ is the number of samples.} are estimated to be $(2.67 \pm 0.27)\%$ for CB24; $(2.10 \pm 0.19)\%$ for CB27 and $(3.11 \pm 0.28)\%$ for CB188,  respectively. The mean orientation of polarization position angle ($<\theta>$) with the standard error are estimated to be $(142.8 \pm 5.7)^\circ$ for CB24; $(145.5 \pm 3.7)^\circ$ for CB27; and $(98.5 \pm 2.3)^\circ$ for CB188, respectively.

\begin{figure}[htb]
	\vspace{-0.5cm}
\begin{center}

	\includegraphics[width=\columnwidth]{{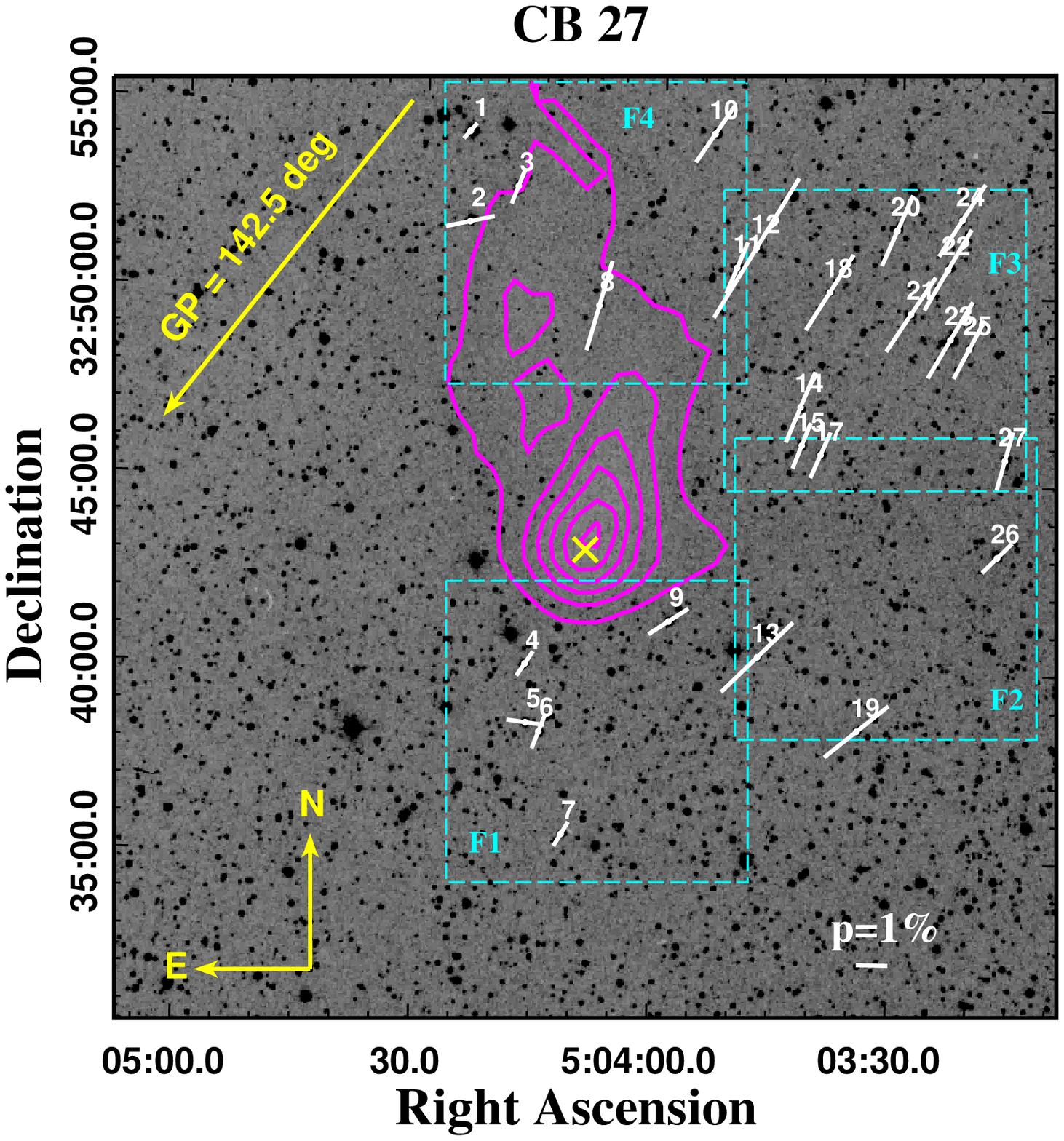}}
	\vspace{-1cm}
        \caption{Polarization map of CB27: White solid lines represent the polarization vectors of the background field stars plotted on a DSS image of the globule CB27 (25$^\prime\times25^\prime$). At the bottom right corner, a vector of 1$\%$ polarization is shown for reference. The vector at the top left corner shows the orientation of the GP ($\theta_{GP}=142.5^\circ$). The cross mark shows the center of the globule. The dashed rectangular boxes of dimension 8$^\prime\times8^\prime$ show the fields of observation (details are given in Table \ref{t}) of the cloud. Also, the contours extracted from \textit{Herschel SPIRE} $500\mu m$ dust continuum emission map in the range of 18 to 74 mJy beam$^{-1}$ with an increasing step size of 8 mJy beam$^{-1}$ are plotted (magenta) over the polarization map.}
    \label{fig2}
    \end{center}
\end{figure}



\begin{table*}
\caption{Polarimetric results of 20 field stars towards CB24. The Right Ascension (RA) and Declination (DEC) of the field stars are given in Column--2 and Column--3, column-4 and 5 represent the degree of linear polarization ($p$) and position angle of polarization ($\theta$), column-6 gives the distance (d in pc) to the individual field stars collected from Gaia EDR3 database. The star having distance more than the distance of the cloud CB24 ($\sim$ 360 pc) is considered to be background to the cloud and is shown in column-7.}
\begin{center}
\begin{tabular}{ccccclc}
\hline
\hline
Star ID & RA(2000)  & Dec(2000)& $p$ $\pm$ $e_p$  & $\theta$ $\pm$ $e_{\theta}$ &  $d$ $\pm$ $e_d$ & Background star  \\
   & (h m s) & ($^\circ$ $^\prime$ $^{\prime\prime}$) & ($\%$) & ($^\circ$) & (pc) & (yes/no) \\
 \hline
1	&	4 59 17.76  & 52 08 31  	&	3.85	$\pm$	0.46	&	146.8	$\pm$	3.4	&	1739	$\pm$	70	&	yes	\\
2	&	 4 59 12.72  & 52 07 33  	&	1.99	$\pm$	0.29	&	86.0	$\pm$	4.2	&	726	    $\pm$	8	&	yes	\\
3	&	 4 59 11.28  & 52 08 09  	&	1.66	$\pm$	0.49	&	98.3	$\pm$	8.4	&	700	    $\pm$	10	&	yes	\\
4	&	 4 59 11.04  & 52 17 34  	&	1.52	$\pm$	0.39	&	158.5	$\pm$	7.3	&	643	    $\pm$	7	&	yes	\\
5	&	 4 59 06.24  & 52 18 21  	&	2.31	$\pm$	0.40	&	161.5	$\pm$	4.9	&	361	    $\pm$	2	&	yes	\\
6	&	 4 59 03.36  & 52 08 09  	&	2.38	$\pm$	0.54	&	143.4	$\pm$	6.5	&	3702	$\pm$	307	&	yes	\\
7	&	 4 58 59.76  & 52 19 15  	&	1.91	$\pm$	0.20	&	159.9	$\pm$	3.0	&	297	    $\pm$	2	&	no	\\
8	&	 4 58 48.72  & 52 20 20  	&	1.39	$\pm$	0.39	&	145.8	$\pm$	8.0	&	817	    $\pm$	38	&	yes	\\
9	&	 4 58 47.28  & 52 12 43  	&	2.24	$\pm$	0.74	&	137.5	$\pm$	9.5	&	2330	$\pm$	164	&	yes	\\
10	&	 4 58 34.80 & 52 19 33  	&	2.00	$\pm$	0.35	&	159.5	$\pm$	5.0	&	501	    $\pm$	14	&	yes	\\
11	&	 4 58 32.16 & 52 18 18  	&	3.54	$\pm$	0.62	&	150.0	$\pm$	5.0	&	~~~~NA	   	&	$-$	\\
12	&	 4 58 27.60 & 52 17 13  	&	2.03	$\pm$	0.25	&	169.1	$\pm$	3.6	&	324	    $\pm$	1	&	no	\\
13	&	 4 58 23.28 & 52 09 39  	&	1.22	$\pm$	0.21	&	125.4	$\pm$	4.8	&	~~~~NA	    	&	$-$	\\
14	&	 4 58 22.80 & 52 16 37  	&	1.76	$\pm$	0.47	&	162.1	$\pm$	7.6	&	334	    $\pm$	2	&	no	\\
15	&	 4 58 21.12 & 52 17 34  	&	3.85	$\pm$	0.66	&	147.1	$\pm$	4.9	&	3891	$\pm$	269	&	yes	\\
16	&	 4 58 18.24 & 52 17 45  	&	1.58	$\pm$	0.29	&	153.6	$\pm$	5.3	&	617	    $\pm$	6	&	yes	\\
17	&	 4 58 12.00 & 52 11 24  	&	3.24	$\pm$	0.35	&	165.0	$\pm$	3.1	&	510	    $\pm$	4	&	yes	\\
18	&	 4 58 10.32 & 52 16 15 	    &	4.01	$\pm$	0.35	&	143.5	$\pm$	2.5	&	3165	$\pm$	226	&	yes	\\
19	&	 4 58 09.36 & 52 18 43 	    &	3.89	$\pm$	1.14	&	145.8	$\pm$	8.4	&	3942	$\pm$	486	&	yes	\\
20	&	 4 58 06.48 & 52 18 36 	    &	4.13	$\pm$	0.84	&	149.7	$\pm$	5.8	&	4498	$\pm$	1354	&	yes	\\

 \hline
\end{tabular}
\end{center}
 \label{t3}
\end{table*}


\begin{figure*}

\begin{center}
\vspace{-0.5cm}

\includegraphics[width=\linewidth]{{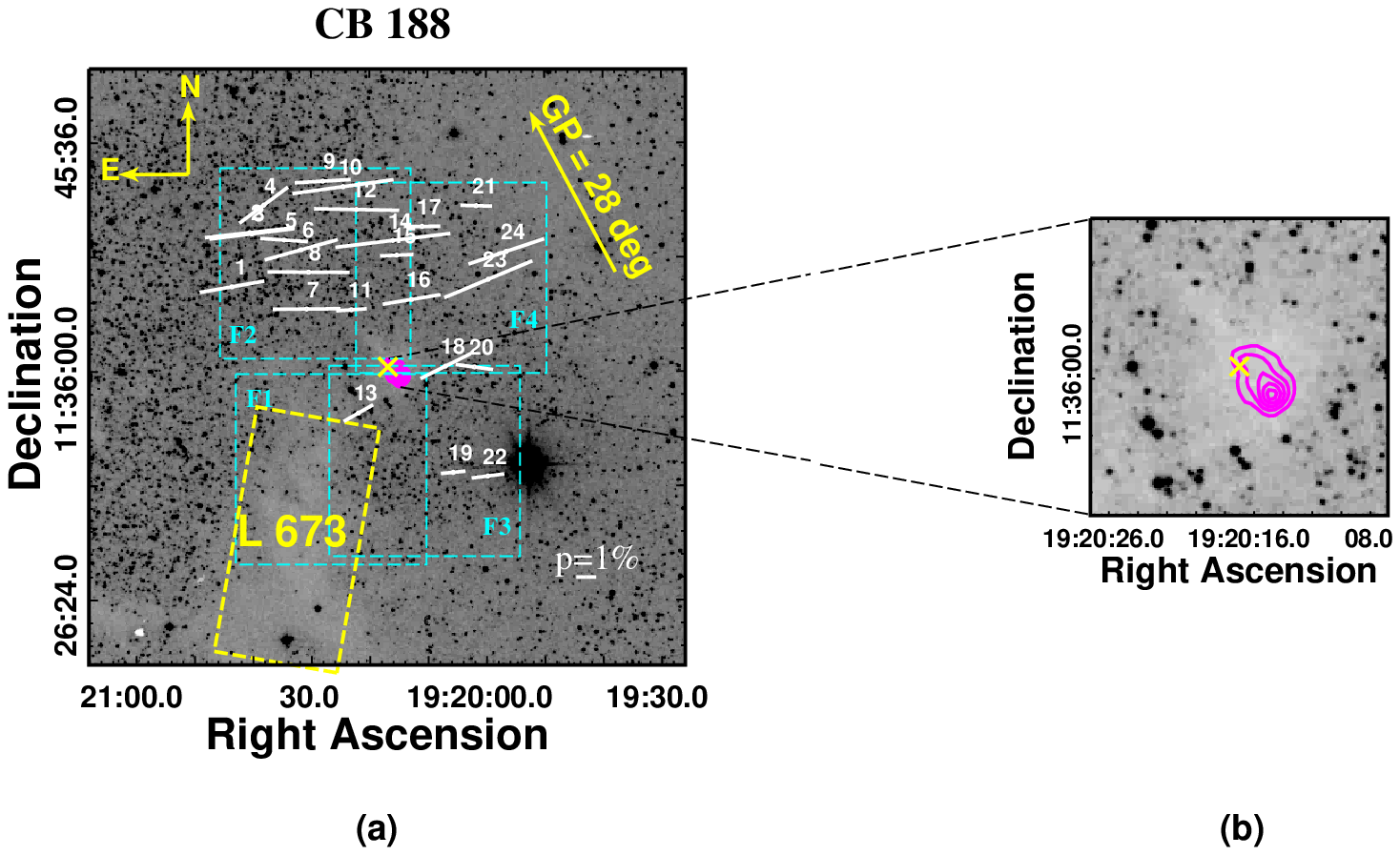}}
\vspace{-1cm}
    \caption{(a) Polarization map of CB188: White solid lines represent the polarization vectors of the background field stars plotted on a DSS image of the globule CB188 (25$^\prime\times25^\prime$). At the bottom right corner, a vector of 1$\%$ polarization is shown for reference. The vector at the top right corner shows the orientation of the GP ($\theta_{GP}=28^\circ$). The cross mark shows the center of the globule. The dashed rectangular boxes of dimension 8$^\prime\times8^\prime$ show the fields of observation (details are given in Table \ref{t}) of the cloud. Also, the contours extracted from \textit{SCUBA} $850\mu m$ dust continuum emission map in the range of 112 to 336 mJy beam$^{-1}$ with an increasing step size of 54 mJy beam$^{-1}$ are plotted (magenta) over the polarization map. The dotted rectangle on the lower region represents the Lynd's cloud L673, which is physically associated with the field of CB188 (see \S \ref{s2.2.2} for details). (b) The zoomed-in view of the central region for better visual of the contours extracted from \textit{SCUBA} $850\mu m$ dust continuum emission map are plotted over a 5$^\prime\times 5^\prime$ DSS image of the Globule CB188.}
    \label{fig3}
    \end{center}
\end{figure*}



\begin{table*}
\caption{Polarimetric results of 27 field stars in CB27. The Right Ascension (RA) and Declination (DEC) of the field stars are given in Column--2 and Column--3, column-4 and 5 represent $p$ and $\theta$ and column-6 gives the distance (d in pc) to the individual field stars collected from Gaia EDR3 database. The star having distance more than the distance of the cloud CB24 ($\sim$ 140 pc) is considered to be background to the cloud and is shown in column-7.}
\begin{center}
\begin{tabular}{ccccclc}
\hline
\hline
   Star ID & RA(2000)  & Dec(2000)& $p$ $\pm$ $e_p$  & $\theta$ $\pm$ $e_{\theta}$ &  $d$ $\pm$ $e_d$ & Background star  \\
   & (h m s) & ($^\circ$ $^\prime$ $^{\prime\prime}$) & ($\%$) & ($^\circ$) & (pc) & (yes/no) \\
 \hline
										
1	&	05 04 24.72 & 32 54 10  	&	0.64	$\pm$	0.19	&	139.8	$\pm$	8.6	&	1255	$\pm$	38	&	yes	\\
2	&	05 04 24.48 & 32 51 46  	&	1.58	$\pm$	0.22	&	103.0	$\pm$	4.0	&	316	$\pm$	5	&	yes	\\
3	&	05 04 18.48 & 32 52 44  	&	1.20	$\pm$	0.34	&	159.5	$\pm$	8.1	&	937	$\pm$	15	&	yes	\\
4	&	05 04 16.32 & 32 40 04  	&	0.93	$\pm$	0.21	&	147.0	$\pm$	6.4	&	2829	$\pm$	168	&	yes	\\
5	&	05 04 16.08 & 32 38 31  	&	1.20	$\pm$	0.31	&	83.1	$\pm$	7.3	&	2266	$\pm$	406	&	yes	\\
6	&	05 04 14.40 & 32 38 16  	&	1.14	$\pm$	0.25	&	159.5	$\pm$	6.4	&	1337	$\pm$	30	&	yes	\\
7	&	05 04 11.28 & 32 35 34  	&	0.89	$\pm$	0.21	&	150.0	$\pm$	6.9	&	1409	$\pm$	35	&	yes	\\
8	&	05 04 07.92 & 32 49 37  	&	2.96	$\pm$	0.57	&	165.1	$\pm$	5.5	&	2286	$\pm$	184	&	yes	\\
9	&	05 03 58.32 & 32 41 16  	&	1.51	$\pm$	0.28	&	123.9	$\pm$	5.4	&	1293	$\pm$	28	&	yes	\\
10	&	05 03 53.76 & 32 54 14  	&	2.18	$\pm$	0.22	&	146.7	$\pm$	2.9	&	1364	$\pm$	32	&	yes	\\
11	&	05 03 50.64 & 32 50 42  	&	1.76	$\pm$	0.47	&	158.4	$\pm$	7.6	&	363	$\pm$	2	&	yes	\\
12	&	05 03 48.24 & 32 51 14  	&	5.20	$\pm$	1.10	&	149.9	$\pm$	6.0	&	1829	$\pm$	117	&	yes	\\
13	&	05 03 47.04 & 32 40 22  	&	3.18	$\pm$	0.35	&	135.4	$\pm$	3.1	&	766	$\pm$	18	&	yes	\\
14	&	05 03 42.24 & 32 47 02  	&	2.43	$\pm$	0.30	&	158.5	$\pm$	3.5	&	1526	$\pm$	63	&	yes	\\
15	&	05 03 42.00 & 32 46 01  	&	1.56	$\pm$	0.28	&	159.3	$\pm$	5.1	&	424	$\pm$	3	&	yes	\\
16	&	05 03 39.84 & 32 48 10  	&	2.76	$\pm$	0.68	&	142.0	$\pm$	7.0	&	~~~~NA			&	--	\\
17	&	05 03 39.60 & 32 45 46  	&	1.59	$\pm$	0.29	&	157.2	$\pm$	5.2	&	1077	$\pm$	19	&	yes	\\
18	&	05 03 38.88 & 32 50 06  	&	2.83	$\pm$	0.49	&	148.8	$\pm$	4.9	&	2826	$\pm$	166	&	yes	\\
19	&	05 03 34.32 & 32 38 27  	&	2.61	$\pm$	0.37	&	129.8	$\pm$	4.1	&	5155	$\pm$	492	&	yes	\\
20	&	05 03 30.48 & 32 51 46  	&	2.45	$\pm$	0.58	&	157.4	$\pm$	6.8	&	655	$\pm$	9	&	yes	\\
21	&	05 03 28.56 & 32 49 33  	&	2.84	$\pm$	0.72	&	147.5	$\pm$	7.2	&	5176	$\pm$	721	&	yes	\\
22	&	05 03 24.00 & 32 50 45  	&	2.97	$\pm$	0.75	&	150.8	$\pm$	7.3	&	5288	$\pm$	590	&	yes	\\
23	&	05 03 23.52 & 32 48 54  	&	2.82	$\pm$	0.34	&	150.7	$\pm$	3.4	&	1860	$\pm$	93	&	yes	\\
24	&	05 03 22.32 & 32 52 04  	&	2.72	$\pm$	0.35	&	148.3	$\pm$	3.7	&	644	$\pm$	10	&	yes	\\
25	&	05 03 21.12 & 32 48 39  	&	2.10	$\pm$	0.27	&	153.1	$\pm$	3.7	&	710	$\pm$	13	&	yes	\\
26	&	05 03 17.04 & 32 43 08  	&	1.33	$\pm$	0.24	&	135.6	$\pm$	5.1	&	1586	$\pm$	69	&	yes	\\
27	&	05 03 16.32 & 32 45 43  	&	1.92	$\pm$	0.21	&	165.4	$\pm$	3.1	&	711	$\pm$	9	&	yes	\\

\hline
\end{tabular}
\end{center}
 \label{t4}
\end{table*}



\begin{table*}
\caption{Plarimetric results of 24 field stars in CB188. The Right Ascension (RA) and Declination (DEC) of the field stars are given in Column--2 and Column--3, column-4 and 5 represent $p$ and $\theta$ and column-6 gives the distance (d in pc) to the individual field stars collected from Gaia EDR3 database. The star having distance more than the distance of the cloud CB24 ($\sim$ 300 pc) is considered to be background to the cloud and is shown in column-7.}
\begin{center}
\begin{tabular}{ccccclc}
\hline
\hline
Star ID & RA(2000)  & Dec(2000)& $p$ $\pm$ $e_p$  & $\theta$ $\pm$ $e_{\theta}$ &  $d$ $\pm$ $e_d$ & Background star \\
   & (h m s) & ($^\circ$ $^\prime$ $^{\prime\prime}$) & ($\%$) & ($^\circ$) & (pc) & (yes/no)\\
 \hline
1	&	19 20 43.57  & 11 39 34 	&	3.30	$\pm$	1.02	&	100.0	$\pm$	8.9	&	2919	$\pm$	167	&	yes	\\
2	&	19 20 40.52  & 11 41 50 	&	4.53	$\pm$	1.16	&	96.3	$\pm$	7.4	&	2520	$\pm$	164	&	yes	\\
3	&	19 20 40.15  & 11 41 47 	&	4.30	$\pm$	1.18	&	96.9	$\pm$	7.9	&	2229	$\pm$	136	&	yes	\\
4	&	19 20 38.12  & 11 42 59  	&	3.04	$\pm$	1.01	&	126.6	$\pm$	9.5	&	666	$\pm$	8	&	yes	\\
5	&	19 20 34.64  & 11 41 31  	&	2.40	$\pm$	0.39	&	86.1	$\pm$	4.6	&	1166	$\pm$	17	&	yes	\\
6	&	19 20 31.70  & 11 41 07  	&	3.81	$\pm$	1.22	&	105.5	$\pm$	9.2	&	2704	$\pm$	184	&	yes	\\
7	&	19 20 30.77  & 11 38 37  	&	3.42	$\pm$	0.68	&	90.4	$\pm$	5.7	&	2498	$\pm$	166	&	yes	\\
8	&	19 20 30.43  & 11 40 10  	&	4.13	$\pm$	1.29	&	89.6	$\pm$	8.9	&	1448	$\pm$	46	&	yes	\\
9	&	19 20 28.00  & 11 43 59  	&	2.82	$\pm$	0.40	&	93.3	$\pm$	4.1	&	1016	$\pm$	18	&	yes	\\
10	&	19 20 24.52  & 11 43 45  	&	5.14	$\pm$	1.45	&	97.8	$\pm$	8.1	&	1462	$\pm$	50	&	yes	\\
11	&	19 20 23.11  & 11 38 35 	&	1.54	$\pm$	0.51	&	93.9	$\pm$	9.5	&	394	$\pm$	3	&	yes	\\
12	&	19 20 22.20  & 11 42 48  	&	4.29	$\pm$	1.21	&	88.8	$\pm$	8.1	&	1453	$\pm$	56	&	yes	\\
13	&	19 20 21.95  & 11 34 15  	&	1.69	$\pm$	0.47	&	119.5	$\pm$	7.9	&	2661	$\pm$	214	&	yes	\\
14	&	19 20 16.00  & 11 41 30  	&	5.82	$\pm$	0.37	&	96.8	$\pm$	1.8	&	2938	$\pm$	298	&	yes	\\
15	&	19 20 15.34  & 11 40 53  	&	1.68	$\pm$	0.53	&	92.5	$\pm$	9.0	&	857	$\pm$	11	&	yes	\\
16	&	19 20 12.79  & 11 39 01  	&	2.94	$\pm$	0.98	&	99.4	$\pm$	9.5	&	929	$\pm$	33	&	yes	\\
17	&	19 20 10.97  & 11 42 05  	&	1.83	$\pm$	0.61	&	90.4	$\pm$	9.6	&	903	$\pm$	15	&	yes	\\
18	&	19 20 07.00  & 11 36 13  	&	2.81	$\pm$	0.93	&	116.4	$\pm$	9.5	&	717	$\pm$	18	&	yes	\\
19	&	19 20 05.72  & 11 31 45  	&	1.24	$\pm$	0.41	&	95.3	$\pm$	9.5	&	869	$\pm$	13	&	yes	\\
20	&	19 20 02.06  & 11 36 11  	&	1.89	$\pm$	0.32	&	82.0	$\pm$	4.9	&	1072	$\pm$	22	&	yes	\\
21	&	19 20 01.69  & 11 42 57  	&	1.58	$\pm$	0.52	&	88.3	$\pm$	9.4	&	838	$\pm$	12	&	yes	\\
22	&	19 19 59.76  & 11 31 36  	&	1.65	$\pm$	0.55	&	97.1	$\pm$	9.5	&	838	$\pm$	11	&	yes	\\
23	&	19 19 59.64  & 11 39 52  	&	4.80	$\pm$	1.51	&	112.7	$\pm$	9.0	&	2479	$\pm$	189	&	yes	\\
24	&	19 19 56.55  & 11 41 02  	&	4.01	$\pm$	1.01	&	108.5	$\pm$	7.1	&	1063	$\pm$	25	&	yes	\\

 \hline
\end{tabular}
\end{center}
 \label{t5}
\end{table*}


Using the values of  $p$ and  $\theta$, polarization maps are generated for the three clouds.  The polarization vectors are plotted on a 25$^\prime\times25^\prime$ DSS (Digital Sky Survey) image of CB24, CB27 and CB188 and are presented in  Figure~\ref{fig1}, Figure~\ref{fig2}, and Figure~\ref{fig3}, respectively. The solid lines represent the polarization vectors whose length represents $p$, and the inclination represents $\theta$. The cross mark represents the center of the cloud. At the bottom right corner of each map, a vector of 1$\%$ polarization is shown for reference. The vector at the top corner represents the orientation of the GP ($\theta_{GP}$), which is $142^\circ$ for CB24; $142.5^\circ$ for CB27 and $28^\circ$ for CB188, respectively. The mean value of the position angle of polarization $<\theta>$ represents the orientation of envelope magnetic field $\theta_B^{env}$ of the cloud i.e. $<\theta>$ = $\theta_B^{env}$. Also, on the polarization map of CB27 (Figure~\ref{fig2}), the contours extracted from \textit{Herschel SPIRE} $500\mu m$ dust continuum emission map are plotted (magenta). The thermal dust continuum map is used to understand the density structure of the globule. In Figure~\ref{fig3}, the $SCUBA$\footnote{SCUBA is Submillimetre Common User Bolometer Array to obtain various astronomical objects. The 850$\mu$ and 450$\mu$ square-degree maps from the Fundamental Dataset and the 850$\mu$ maps from the Extended Dataset are available for download from the SCUBA Legacy catalogues repository at the Canadian Astronomical Data Centre (CADC) at: \url{http://www.cadc.hia.nrc.gc.ca/community/scubalegacy}}
$850 \mu m$ dust continuum emissions is overlaid (magenta) on the polarization map of CB188. The region of cloud L673 which is physically associated with the field of CB188 as mentioned in \S \ref{s2.3} is also marked by the yellow dotted rectangle in Figure~\ref{fig3}.

It is evident from Figure~\ref{fig1} and Figure~\ref{fig2} that the polarization vectors of all the field stars are more or less unidirectional and almost aligned along the GP. As can be further noticed from the contours over-plotted on the polarization map of CB27 (Figure~\ref{fig2}) that the polarization vectors are oriented along the direction of the core (extracted from the \textit{SPIRE} data). Also, there are two polarization vectors (\#8 \& \#3) in the range of contours that show parallel orientation with the alignment of the core. The offset between the envelope magnetic field (given by the mean orientation of the polarization vectors) and the orientation of the GP, $\theta_{off}=|\theta_B^{env}-\theta_{GP}|$, which is $0.8^\circ$ in CB24 and $2.9^\circ$ in CB27 respectively. So, the envelope magnetic field orientation is clearly aligned along the GP in both the clouds. A similar trend was observed for cloud CB17 by \citet{choudhury_bok_2019}. In contrast, in the case of CB188, $\theta_{off} = 70.5^\circ$ (Figure~\ref{fig3}) though all the polarization vectors are unidirectional. Also, the envelope magnetic field orientation is different from the alignment of the $850\mu m$ dust emission contours. Thus, it can be inferred that the orientation of the envelope magnetic field in CB188 is not parallel with the GP, unlike CB24 and CB27. To build a basis for statistically relevant conclusions about the relative orientations of magnetic field traced in the envelope region of the globules with respect to the GP, we include polarimetric data of 14 further low galactic latitude ($-10^\circ < b < 10^\circ$) clouds from the literature. The corresponding results are summarised in Table~\ref{t6}.

\begin{figure}
\begin{center}
\includegraphics[totalheight=0.45\textheight]{{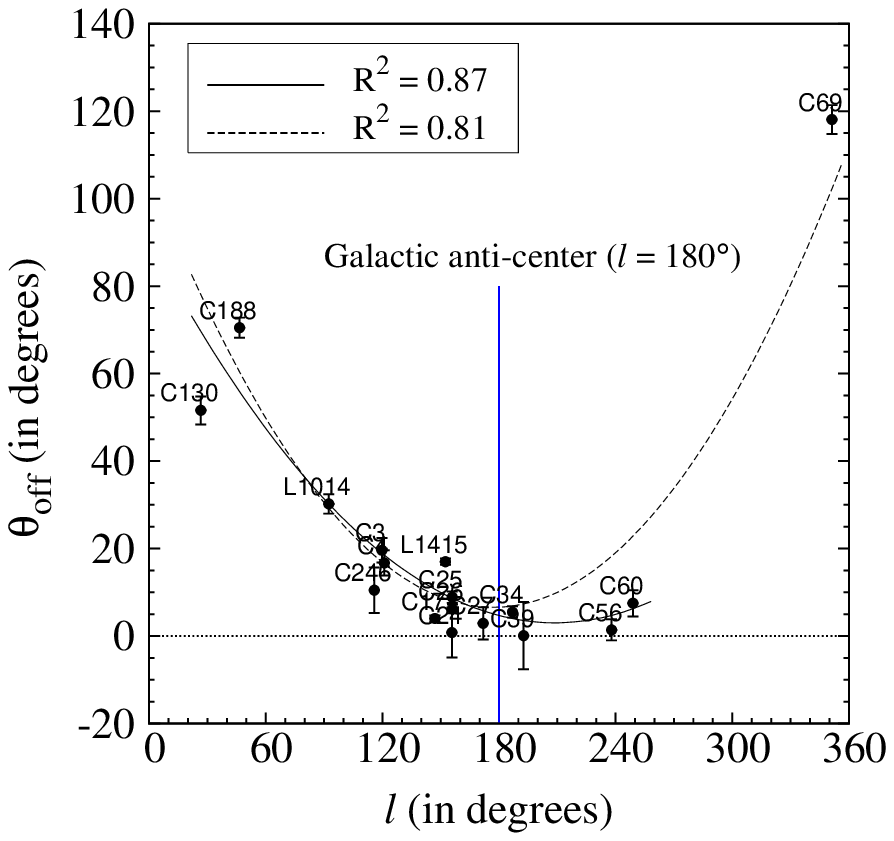}}
 \caption{
 The variation of the alignment of envelope magnetic field of star forming clouds along the GP, $\theta_{off}=|\theta_B^{env}-\theta_{GP}|$ (along Y-axis) with their galactic longitude $l$ (along X-axis) in case of low latitude clouds ($-10^\circ < b < 10^\circ$). ``C" stands for CB cloud and ``L" stands for Lynd's cloud. Also, note that the errors associated with the $\theta_B^{env}$ are the standard error of the mean.
 }
  \label{fig4}
 \end{center}

\end{figure}



\begin{figure}
\centering
\vspace{-0.5cm}
 \includegraphics[totalheight=0.35\textheight, trim=2.3cm 0.5cm 0cm 0.6cm, clip=true, angle=0]{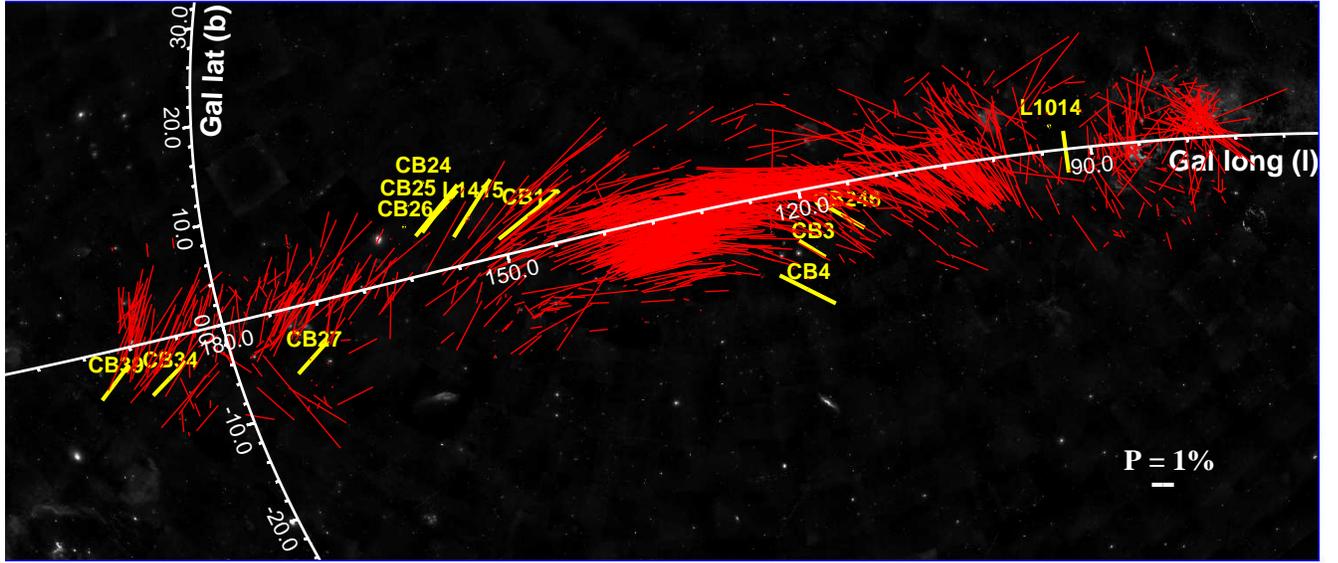}
 \caption{The stellar polarization vectors obtained from Heiles catalogue (shown by red line) over the ranges $-10^\circ < b < 10^\circ$ and $88^\circ< l <195^\circ$ are plotted along with the mean degree of polarization and position angle of polarization of the stars background to the clouds (taken from Table~\ref{t6} and shown by yellow lines) viz. CB3, CB4, CB17, CB24, CB25, CB26, CB27, CB34, CB39, CB246, L1415 and L1014. These clouds are situated towards the region of GAC except for L1014 which is situated in the region towards GC. A reference polarization vector of 1$\%$ polarization is shown on the bottom right corner.}
\label{fig5}
\end{figure}



\begin{figure}
\begin{center}
\includegraphics[totalheight=0.4\textheight]{{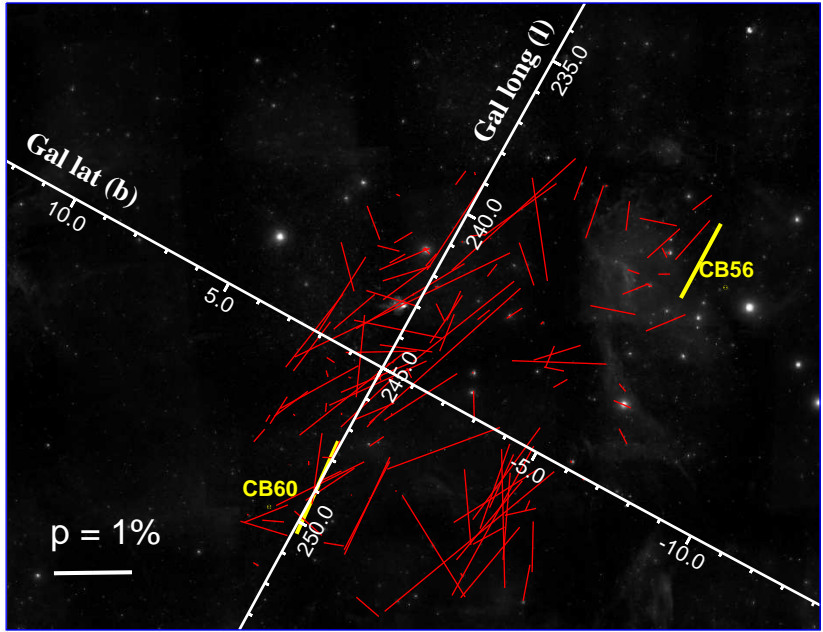}}

 \caption{The stellar polarization vectors obtained from Heiles catalogue (shown by red line) over the ranges $-10^\circ < b < 10^\circ$ and $235^\circ< l <250^\circ$ are plotted along with the mean degree of polarization and position angle of polarization of the stars background to the two clouds (taken from Table~\ref{t6} and shown by yellow lines) viz. CB56 and CB60 situated towards the region of GAC. A reference polarization vector of 1$\%$ polarization is shown on the bottom left corner.}
  \label{fig6}
 \end{center}

\end{figure}



\begin{figure}
\begin{center}
\includegraphics[totalheight=0.6\textheight, trim=2.5cm 0cm 0cm 0cm, clip=true, angle=0]{{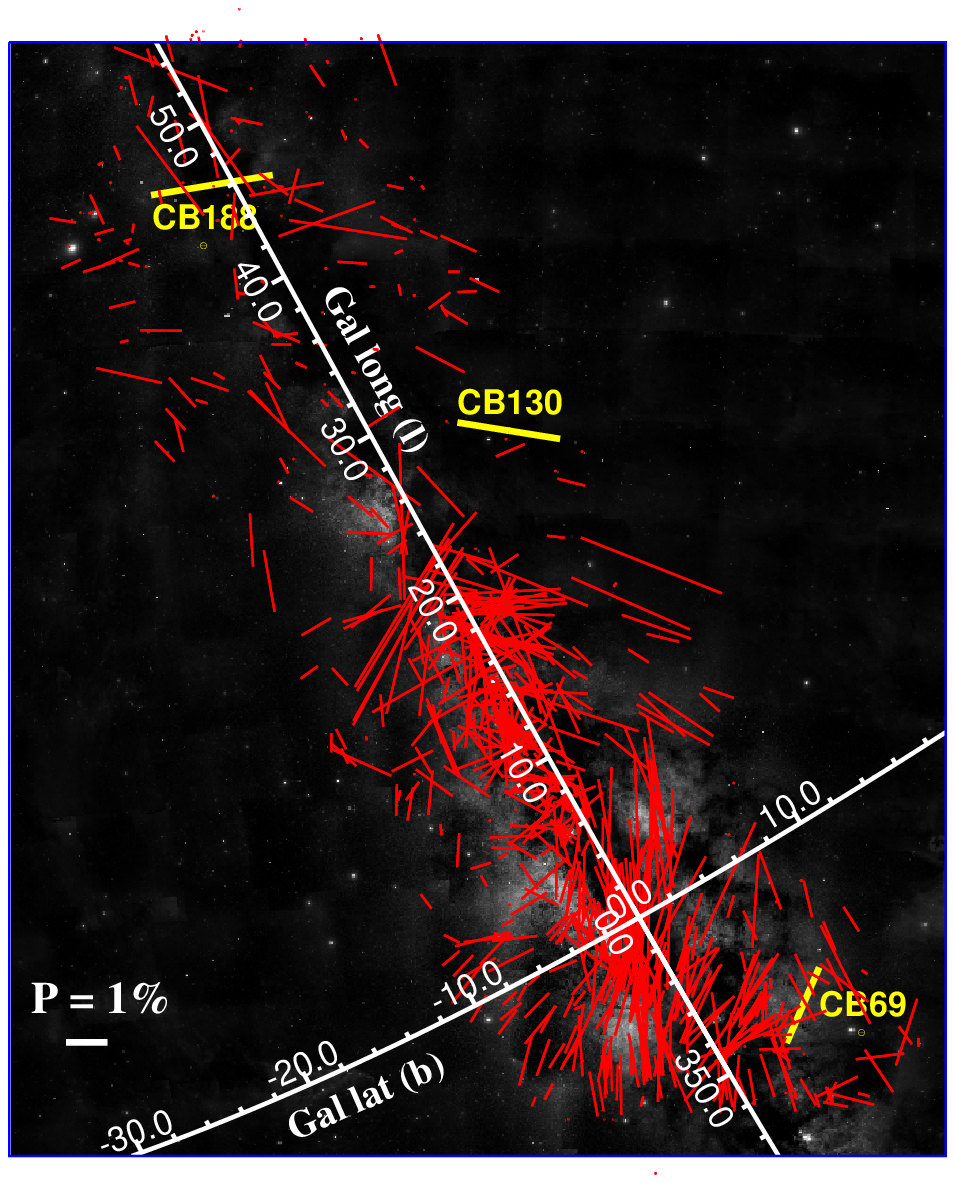}}
 \caption{The stellar polarization vectors obtained from Heiles catalogue (shown by red line) over the ranges $-10^\circ < b < 10^\circ$ and $l>345^\circ$ and $l<50^\circ$ are plotted along with the mean degree of polarization and position angle of polarization of the stars background to the three clouds (taken from Table~\ref{t6} and shown by yellow lines) viz. CB69, CB130 and CB188 situated towards the region of GC. A reference polarization vector of 1$\%$ polarization is shown on the bottom left corner.}
  \label{fig7}
 \end{center}

\end{figure}

\section{Results and Discussion}
\label{s4}

\begin{table*}

\caption{Parameters related to the target globules obtained from our study as well from literature: Cloud ID, galactic longitude ($l$), galactic latitude ($b$), mean value of degree of polarization ($<p>$), position angle of GP ($\theta_{GP}$), position angle of envelope magnetic field ($\theta_{B}^{env}$), offset ($\theta_{off}=|\theta_B^{env}-\theta_{GP}|$), FWHM line width ($\Delta V$), uncertainty associated  with $\Delta V$  and position angle of core magnetic field ($\theta_{B}^{core}$).}
\begin{center}
\scriptsize{
\begin{tabular}{r r r c r r c l c c}
\hline
\hline
 ID  & $l$ & $b$ & $<p>$ & $\theta_{GP}$ $\pm$ $e_{\theta_{GP}}$   &   $\theta_{B}^{env}$ $\pm$ $e_{\theta_{B}^{env}}$ & $\theta_{off}$ $\pm$ $e_{\theta_{off}}$ & $\Delta$ V $\pm$ $e_{\Delta V}$ & $e_{\Delta V}$ associated   & $\theta_{B}^{core}$ \\
	&     &  &  &  && (FWHM) & with $\Delta V$ are: &      \\	
	&  ($^\circ$)    & ($^\circ$) & ($\%$) & ($^\circ$) & ($^\circ$)  & ($^\circ$)& (kms$^{-1}$) && ($^\circ$)    \\
 \hline

CB3 	&	119.8	&	$-$6.03	&	1.41	&	85.0$\pm$0.03	&	65.4$\pm$2.8	&	19.6$\pm$2.8	&	1.60$\pm$0.03$^{a}$	& S.E.$^{*}$    &	69.0$^{f}$  	\\
CB4	&	121.03	&	$-$9.96	&	2.84	&	87.2$\pm$0.03	&	70.6$\pm$2.9	&	16.7$\pm$2.9	&	0.51$\pm$0.01$^{c}$ 	& S.E.        &	$-$	 \\
CB17	&	147.02	&	3.39	&	3.52	    &	132.0$\pm$0.04	&	136.0$\pm$0.7	&	4.0$\pm$0.7	    &	0.97$\pm$0.03$^{c}$	& S.E.          &	44.0$^{g}$	\\
CB24	&	155.76	&	5.9	&	2.67	    &	142.0$\pm$0.04	&	142.8$\pm$5.7	&	0.8$\pm$5.7	    &	0.80$\pm$0.50$^{b}$	& S.D.$^{\dag}$ &	$-$	\\
CB25	&	155.97	&	5.84	&	2.35	    &	142.0$\pm$0.04	&	150.9$\pm$1.3	&	8.9$\pm$1.3	    &	0.70$\pm$0.50$^{b}$	& S.D.          &	$-$	\\
CB26	&	156.05	&	5.99	&	3.00	    &	142.0$\pm$0.04	&	148.2$\pm$1.0	&	  6.2$\pm$0.0	    &	 1.17$\pm$0.02$^{c}$ 	& S.E.          &	25.3$^{h}$	\\
CB27	&	171.82	&	$-$5.18	&	2.10	    &	142.6$\pm$0.04	&	145.5$\pm$3.7	&	2.9$\pm$3.7	    &	0.89$\pm$0.01$^{c}$	& S.E. &	$-$	\\
CB34	&	186.94	&	$-$3.83	&	2.14	    &	148.8$\pm$0.04	&	143.3$\pm$1.3	&	5.5$\pm$1.3	    &	1.50$\pm$0.08$^{a}$	& S.E. &	46.7 for Core1$^{i}$	\\
	&		&		&		&		&		&	&	&		&	90.4 for Core2$^{i}$	\\
CB39	&	192.63	&	$-$3.04	&	1.95	    &	150.4$\pm$0.03	&	150.3$\pm$7.7	&	0.1$\pm$7.7	    &	2.05$\pm$0.50$^{b}$	& S.D.          &	$-$	\\
CB56	&	237.9	&	$-$6.45	&	1.08	    &	152.3$\pm$0.02	&	150.9$\pm$2.4	&	1.4$\pm$2.4	    &	1.44$\pm$0.50$^{b}$	& S.D.          &	$-$	\\
CB60	&	248.89	&	$-$0.01	&	1.30	    &	147.7$\pm$0.01	&	155.2$\pm$3.0	&	7.5$\pm$3.0	    &	1.82$\pm$0.40$^{b}$	& S.D.          &	$-$	\\
CB69	&	351.23	&	5.14	&	2.00	&	37.7$\pm$0.04	&	155.8$\pm$3.3	&	118.1$\pm$3.3	&	2.35$\pm$0.50$^{b}$	& S.D.          &	$-$	\\
CB130	&	26.61	&	6.65	&	2.53	&	28.4$\pm$0.03	&	80.0$\pm$3.2	&	51.6$\pm$3.2	&	4.20$^{c}$ $\pm$ $-$	 & $-$        &	$-$	\\
CB188	&	46.53	&	$-$1.02	&	3.11	&	28.0$\pm$0.01	&	98.5$\pm$2.3	&	70.5$\pm$2.3	&	4.40$\pm$1.1$^{b}$	& S.D.          &	$-$	\\
CB246	&	115.84	&	$-$3.54	&	1.92	&	77.9$\pm$0.03	&	67.4$\pm$5.2	&	10.5$\pm$5.2	&	1.62$\pm$0.50$^{b}$	& S.D.          &	$-$	\\
L1014	&	92.45	&	$-$0.12	&	1.90	&	45.2$\pm$0.02	&	15.0$\pm$2.2	&	30.2$\pm$2.2	&	2.26$\pm$0.05$^{d}$ 	& S.E. &	$-$	\\
L1415	&	152.41	&	5.27	&	3.10	&	138.0$\pm$0.04	&	155.0$\pm$0.7	&	17.0$\pm$0.7	&	1.65$\pm$0.02$^{e}$	& S.E. &     	$-$	\\

\hline
\end{tabular}
}
\end{center}
\vspace{.05cm}
$^{(a)}$\citet{wang_collapse_1995}
$^{(b)}$\citet{clemens_bok_1991}
$^{(c)}$\citet{lippok_gas-phase_2013}
$^{(d)}$\citet{crapsi_dynamical_2005}
$^{(e)}$\citet{soam_probing_2017}
$^{(f)}$\citet{ward-thompson_optical_2009}
$^{(g)}$\citet{choudhury_bok_2019}
$^{(h)}$\citet{henning_measurements_2001}
$^{(i)}$\citet{das_magnetic_2016}

$^{*}$S.E.: Standard error of the mean \&
$^{\dag}$S.D: Standard deviation
 \label{t6}
\end{table*}

 In this section, we discuss the results of the presented polarization measurements. In Table~\ref{t6}, we present the angular offset in the orientation of envelope magnetic field ($\theta_{B}^{env}$, traced through optical polarimetry) with the orientation of GP ($\theta_{GP}$). The uncertainties in $\theta_B^{env}$ considered here are the standard error of the mean. Note that, due to the unavailability of the orientation of core-scale magnetic field ($\theta_{B}^{core}$) for the majority of the clouds, it is not possible to estimate the morphology of the core-magnetic field of the clouds. The interpretations based on the results obtained are discussed in the following subsections.

\subsection{Relative orientation between the magnetic field and the Galactic plane}
\label{s4.1}
 Various studies were carried out to find a correlation between the orientation of envelope magnetic field in molecular clouds with the orientation of the GP \citep[e.g.][]{ sen_imaging_2000, soam_magnetic_2015, das_magnetic_2016, chakraborty_study_2016, choudhury_bok_2019}. The magnetic lines of force in the spiral arm of our galaxy are parallel to the arm everywhere \citep{ireland_effect_1961}.
In our previous work, the polarimetric study of CB17 reveals that the projected envelope magnetic field of the globule is oriented along the GP \citep{choudhury_bok_2019}.

Based on the results of 17 clouds presented here, the mean value of position angle of polarization ($\theta_{B}^{env}$) determined for 13 clouds shows that the envelope magnetic field is almost aligned along the position angle of GP ($\theta_{GP}$) (see Figure~\ref{fig4}).
The offset between the orientation of magnetic filed and the GP ($\theta_{off}$) for these 13 clouds are within 20$^\circ$ with an average offset of $7.8^\circ$.
However, for other 4 clouds, a decoupling in the relative orientation between the magnetic field and the GP is observed with
an average offset of $67.6^\circ$. The details are listed in Table~\ref{t6}.

\subsection{Variation in the relative orientation between the magnetic field and the GP with the galactic longitude}
\label{s4.2}
 In this section, we discuss the possible correlation in the relative orientation between the magnetic field and the GP with the galactic longitude.
In Figure~\ref{fig4}, we show the variation in the offset between the orientation of envelope magnetic field and the GP, $\theta_{off}$ ($= |\theta_B^{env}-\theta_{GP}|$) with the galactic longitude ($l$). In this plot, we have considered only the magnitude of the offset and not the sign.

In our sample size, we have adequate data points in longitude range $0^\circ$ to $250^\circ$. A second order polynomial fitting is done with these data points (solid curve) and a strong correlation is observed between $l$ and $\theta_{off}$, with an equation, $\theta_{off} = a_{1}.l^2 - b_{1}.l + c_{1}$, where $a_{1}=0.0020 \pm 0.0006$, $b_{1}=0.8380 \pm 0.1841$, and $c_{1}=90.5358 \pm 14.09$. The fitting is done by including the error (standard error of the mean) in $\theta_{off}$. To test the goodness of the fit, we have estimated the co-efficient of determination ($R^2$) of the best fitted equation. $R^2$ is a key output of regression analysis that may be interpreted as the proportion of the variance in the dependent variable predicted from the independent variable, which lies between 0 and 1. The higher the coefficient, the better is the goodness of fit. In this case, $R^2$ is estimated to be $\approx  0.87$. It is evident from this figure that the offset between the orientation of GP and the envelope magnetic field is considerably low in the regions of $115^\circ<l<250^\circ$ (which corresponds to the region towards galactic anti-center (GAC)). This indicates that the local magnetic of the clouds situated towards the GAC region is oriented along the GP. The significant offset between the $\theta_B^{env}$ and $\theta_{GP}$  observed in the region $l<100^\circ$  (correspond to the region towards the galactic center, (GC)) show that the orientation of the magnetic field in the clouds lying in those regions (CB130, CB188 and L1014) tend to become perpendicular to the GP. Hence, for the clouds located at those longitudes, the envelope magnetic field has a local deflection of its own irrespective of the orientation of GP.

Note that, no CB cloud is located in the region $250^\circ<l<350^\circ$ in the molecular clouds catalogue of \citet{Clemens1988}. However, since there is one single cloud (CB69) available between $350^\circ$ and $360^\circ$ which has a huge offset between the orientation of envelope magnetic field and GP, we have included this cloud in our sample size to observe the best fitting parameters. A second order polynomial fitting is done which is represented by the dashed curve. The curve appears to follow the same trend as in the region $l<100^\circ$, which indicates that, towards the GC region the clouds possess their own local deflection irrespective of the orientation of GP. In this case, $R^2$ is estimated to be $\approx 0.81$ and the fitting equation is $\theta_{off} = a_{2}.l^2 - b_{2}.l + c_{2}$, where $a_{2}=0.0032 \pm 0.0004$, $b_{2}=1.1177 \pm 0.1484$, and $c_{2}=105.559 \pm 13.82$. Although we have made an attempt to fit the equation including CB69, the fitting parameters obtained may not be significant enough due to insufficient data points beyond $l > 250^\circ$.

We compare the polarization of the clouds with the polarization of the stars by \citet{2000AJ....119..923H} to interpret the relative orientation of the local magnetic field and the GP. The stellar polarization catalogues agglomerated by \citet{2000AJ....119..923H} contains polarization data of 9268 stars in the whole sky ($-90^\circ<b<90^\circ$) with $p$ and $\theta$ ranging from $0-12.47\%$ and $0^\circ-180^\circ$. As the region of interest in our study is $-10^\circ<b<10^\circ$, we have extracted 2386 polarization data of stars from the Heiles catalogue within this range. In Figure~\ref{fig5}, Figure~\ref{fig6}, and Figure~\ref{fig7}, the mean degree of polarization ($<p>$) and position angle of polarization vectors ($<\theta>$) of the field stars located in the clouds are overlaid along with the stellar polarization vectors obtained from Heiles catalogue on DSS image.
The red lines represent stellar polarization vectors obtained from the Heiles catalogue and yellow lines represent the mean polarization vectors of the stars background to the clouds (shown by the cloud IDs).
In Table~\ref{t8}, we have presented the mean position angle of stellar polarization vectors ($<\theta_{Heiles}>$) and the standard deviation ($\sigma_{\theta_{Heiles}}$) in the selected longitude range where the studied clouds are located.
The polarization vectors of 1523 stars obtained from Heiles catalogue are plotted in Figure~\ref{fig5} along with $<p>$ and $<\theta>$ of twelve clouds viz., CB3, CB4, CB17, CB24, CB25, CB26, CB27, CB34, CB39, CB246, L1014 and L1415 in the longitude range $88^\circ< l <195^\circ$.
In Figure~\ref{fig6}, 156 stellar polarization vectors are plotted along with $<p>$ and $<\theta>$ of CB56 and CB60, situated in the longitude range $235^\circ< l <250^\circ$ that corresponds to the region of GAC. We have also presented polarization vectors of 724 stars along with $<p>$ and $<\theta>$ of CB69, CB130, and CB188 situated in the longitude range $l>350^\circ$ and $l<60^\circ$ corresponds to the region of GC in Figure~\ref{fig7}. Misalignment among stellar polarization vectors at particular longitude ranges is noted to be strong when $\frac{<\theta_{Heiles}>}{\sigma_{\theta_{Heiles}}}$ $<2$ (see Table~\ref{t8}). As revealed from Figure~\ref{fig5} and Table~\ref{t8}, the observed nine star-forming clouds in the longitude range $110^\circ< l <180^\circ$ ($\frac{<\theta_{Heiles}>}{\sigma_{\theta_{Heiles}}}$ $>2$) are found to be almost aligned with the stellar polarization vectors at their local regions which corresponds towards the region of GAC. A slight randomness is observed in the region $185^\circ< l <195^\circ$ ($\frac{<\theta_{Heiles}>}{\sigma_{\theta_{Heiles}}}$ $=1.67$) where CB34 and CB39 are located. However, in the region around $l\sim90^\circ$ towards GC ($\frac{<\theta_{Heiles}>}{\sigma_{\theta_{Heiles}}}$ $=1.08$), a significant randomness in the alignment of stellar polarization vectors is seen where the cloud L1014 is located. In Figure~\ref{fig6}, almost unidirectional orientation of the stellar polarization vectors is observed in the region $235^\circ< l <250^\circ$ ($\frac{<\theta_{Heiles}>}{\sigma_{\theta_{Heiles}}}$ $\approx2$). However, in Figure~\ref{fig7}, misalignment in the orientation of stellar polarization vectors as well as the star-forming clouds is observed in the regions $345^\circ< l <355^\circ$ and $20^\circ< l <50^\circ$ ($\frac{<\theta_{Heiles}>}{\sigma_{\theta_{Heiles}}}$ $<2$). Thus, a local irregularity is observed towards GC revealed from the stellar polarization data of Heiles, which is found to be consistent with our results.


\begin{table*}

\caption{Heiles polarization data averaged over particular galactic longitude ranges: longitude range over which the mean polarization is estimated ($l-$range), mean position angle of Heiles polarization vectors ($<\theta_{Heiles}>$), standard deviation in the position angle of Heiles polarization ($\sigma_{\theta_{Heiles}}$), number of polarization vectors found (n), position angle of envelope magnetic field averaged over star forming clouds located in the longitude range given in column-1 along with the standard deviation ($<\theta_{B}^{env}>$ $\pm$ $\sigma_{\theta_{B}^{env}}$) and cloud IDs.}
\begin{center}
\scriptsize{
\begin{tabular}{c c c c c c c}
\hline
\hline
 $l$  &  $<\theta_{Heiles}>$ & $\sigma_{\theta_{Heiles}}$   & $<\theta_{Heiles}>$/$\sigma_{\theta_{Heiles}}$ & n$^{\dag}$ &   $<\theta_{B}^{env}>$ $\pm$ $\sigma_{\theta_{B}^{env}}$ & Cloud IDs \\
  ($^\circ$)    & ($^\circ$) & ($^\circ$) & &  & ($^\circ$)&   \\
 \hline

20--30	&	89.47	&	52.39	&	1.71	&	43	&	80 $\pm$ 3.20	&	CB130	\\
40--50	&	72.77	&	48.01	&	1.52	&	65	&	98.5 $\pm$ 2.30	&	CB188	\\
88--100	&	53.90	&	49.81	&	1.08	&	89	&	15.0 $\pm$ 2.2	&	L1014	\\
110--122	&	77.01	&	27.69	&	2.78	&	129	&	67.8 $\pm$ 2.58	&	CB3, CB4, CB246	\\
147--157	&	130.49	&	29.70	&	4.39	&	81	&	147.10 $\pm$ 7.14	&	CB17, CB24, CB25, CB26, L1415	\\
167--177	&	122.13	&	54.91	&	2.22	&	92	&	145.5 $\pm$ 3.7	&	CB27	\\
185--195	&	113.72	&	68.26	&	1.67	&	58	&	146.79 $\pm$ 4.93	&	CB34, CB39	\\
235--250	&	103.83	&	53.13	&	1.95	&	156	&	153.05 $\pm$ 2.98	&	CB56, CB60	\\
345--355	&	79.06	&	67.66	&	1.17	&	177	&	155.8 $\pm$ 3.3	&	CB69	\\

\hline
\end{tabular}
}

$^{\dag}$ Number of Heiles polarization vectors present in the longitude range given in column-1
\end{center}

 \label{t8}
\end{table*}


Based on the theory given by \citet{davis_polarization_1951}, \citet{ireland_effect_1961} found that polarization effect tends to attain maximum intensity in galactic longitudes close to $102^\circ$. At such longitude, the direction of polarization is close to being parallel to the plane of the galaxy. \citet{ireland_effect_1961} also found that, for $70^\circ  < l < 130^\circ$, polarization is high and the magnetic field lines are oriented along to the GP in Orion arm. However, for $170^\circ < l < 220^\circ$, the polarization is much weaker, and there is a marked tendency for a few polarization to be normal to the GP. \citet{berkhuijsen1964linear} while studying the linear polarization of the galactic background observed that there is a homogeneous magnetic field parallel to the GP around $l=140^\circ$, $b=6^\circ$. Generally, the magnetic field lines of the Milky Way galaxy are found to follow the orientation of the spiral arms \citep{han_pulsar_2006}. \citet{fosalba_statistical_2002} observed a net alignment of the magnetic field with Galactic structures on large scales. \citet{beck2013magnetic} found evidence of turbulence in polarized intensity towards the inner Galaxy ($270^\circ < l < 90^\circ$, $|b| < 30^\circ$).
The region towards the GC holds most uniform fields of up to milligauss strength that are oriented normal to the plane \citep{alfaro_magnetic_2004}. Thus, the results obtained from this study is in good agreement with the previous studies of the homogeneous magnetic field parallel to the GP for a certain longitude range as discussed above.

\subsection{The effect of turbulence on the cloud}
\label{s4.3}
 The GC is considered to have higher turbulence, indicating high activities in star forming regions \citep{boldyrev_turbulent_2006}. So, the observed misalignment in the orientation between the envelope magnetic field and the GP towards the GC led us to study the effect of turbulence. The $^{12}$CO line-width or velocity dispersion values are considered to be a good measure of turbulence in molecular clouds. We have listed the $^{12}$CO line-width ($\Delta V$ km s$^{-1}$ in column-7 of Table~\ref{t6}) taken from \citet{wang_collapse_1995}, \citet{clemens_bok_1991}, \citet{lippok_gas-phase_2013}, \citet{crapsi_dynamical_2005} and \citet{soam_probing_2017}. The uncertainties associated with $\Delta V$ taken from \citet{clemens_bok_1991} are the dispersion of the distribution and not the standard error of the mean, they provided the dispersion based on three cloud categories viz. Group-A (uncertainty=0.5), Group-B (uncertainty=0.4) and Group-C (uncertainty=1.1). In our sample, the clouds CB24, CB25, CB39, CB56, CB69, and CB246 fall into Group-A category; CB60 falls into Group-B and CB188 into Group-C (See Table~3 of \citet{clemens_bok_1991} for details).

It can be seen from Table~\ref{t6} (column-7) that the clouds which show noticeable misalignment between the envelope magnetic field and the orientation of GP ($\theta_{off} >30^\circ$) are seen to have comparatively higher $\Delta V$ ($ > 2$ km s$^{-1}$). Thus, it can be commented that the clouds having higher $\Delta V$, which is an indication of more dynamical activities  within the cloud, are seen to have weaker alignment amongst the polarization vectors. However, most of the clouds with $\Delta V$ $<$ 2 kms$^{-1}$ appears to have low $\theta_{off}$ ($<$ 20$^\circ$), which shows that the polarization vectors of the molecular clouds with less dynamical activities show comparatively better alignment amongst themselves as well as with the orientation of the GP. The clouds having more dynamical activities exhibit randomness in the alignment of polarization vectors. This is because the regions with high turbulent activities are supposed to be warmer and have better grain alignment. However, precise measurements of the turbulence velocity is not available for these clouds preventing us from reaching firm conclusions.

\section{Conclusions}
\begin{enumerate}
\item  We present optical polarimetric analysis of three Bok Globules CB24, CB27 and CB188. The observations were conducted with the 104cm ST in R-band at ARIES, Nainital, India. The mean value of polarization, $<p>$ along with the standard error are found to be $(2.67 \pm 0.27)\%$, $(2.10 \pm 0.19)\%$, and $(3.11 \pm 0.28)\%$ for CB24, CB27, and CB188, respectively. The mean value of polarization position angle, $<\theta>$ with the standard error are estimated to be $(142.8 \pm 5.7)^\circ$, $(145.5 \pm 3.7)^\circ$, and $(98.5 \pm 2.3)^\circ$ for CB24, CB27, and CB188, respectively.

\item As revealed by the imaging polarimetry, we have found that the envelope magnetic field in CB24 and CB27 are aligned along the GP. However, in CB188, the envelope magnetic field is almost normal to the GP. Since all these three clouds are situated close to GP, the dissimilarities in the results motivated us to extend our study for 14 more low galactic latitude clouds, which are available in the literature.

\item Based on the observational evidences discussed in \S \ref{s4}, we may reasonably conclude that the magnetic field has its own local deflection irrespective of the orientation of GP in the clouds which are situated in the region $l<100^\circ$ towards the GC. However, in the region $100^\circ<l<250^\circ$ towards the of GAC, the conventional view of the orientation of magnetic field of the clouds along the GP is observed. This is apparent from the fit obtained between $l$ and $\theta_{off}$ by a second order polynomial equation, $\theta_{off} = a_{1}.l^2 - b_{1}.l + c_{1}$, where $a_{1}=0.0020 \pm 0.0006$, $b_{1}=0.8380 \pm 0.1841$, and $c_{1}=90.5358 \pm 14.09$ with $R^2 = 0.87$ (solid black curve, Figure~\ref{fig4}). However, on inclusion of the single cloud CB69 situated at a galactic longitude of $351.23^\circ$ towards GC, the fitting equation becomes $\theta_{off} = a_{2}.l^2 - b_{2}.l + c_{2}$, where $a_{2}=0.0032 \pm 0.0004$, $b_{2}=1.1177 \pm 0.1484$, and $c_{2}=105.559 \pm 13.82$ with $R^2 = 0.81$ (dashed black curve, Figure~\ref{fig4}).

\item We have compared our results with the stellar polariation data obtained from Heiless catalogue. We have found a misalignment among stellar polarization vectors towards the region of GC where $\frac{<\theta_{Heiles}>}{\sigma_{\theta_{Heiles}}} <2$. In the longitude range $110^\circ< l <180^\circ$ (region towards GAC), almost unidirectional orientation in the stellar polarization vectors is seen, which is also observed in our study. We have also noted a little misalignment in $185^\circ< l <195^\circ$. However, in the longitude range $20^\circ< l <50^\circ$, $l\sim90^\circ$, and $345^\circ< l <355^\circ$ (region towards GC), a strong misalignment in the orientation of the stellar polarization vectors is observed, which are in good agreement with our results.

\item The presence of high turbulent activities towards the GC makes the star forming clouds dynamically more active. Hence, the high turbulence may possibly play a pivotal role on the misalignment between the magnetic field and the GP.
\end{enumerate}







\def\authorcontribution{Author contribution}%

\def\aj{AJ}%
\def\actaa{Acta Astron.}%
\def\araa{ARA\&A}%
\def\apj{ApJ}%
\def\apjl{ApJ}%
\def\apjs{ApJS}%
\def\ao{Appl.~Opt.}%
\def\apss{Ap\&SS}%
\def\aap{A\&A}%
\def\aapr{A\&A~Rev.}%
\def\aaps{A\&AS}%
\def\azh{AZh}%
\def\baas{BAAS}%
\def\bac{Bull. astr. Inst. Czechosl.}%
\def\caa{Chinese Astron. Astrophys.}%
\def\cjaa{Chinese J. Astron. Astrophys.}%
\def\icarus{Icarus}%
\def\jcap{J. Cosmology Astropart. Phys.}%
\def\jrasc{JRASC}%
\def\mnras{MNRAS}%
\def\memras{MmRAS}%
\def\na{New A}%
\def\nar{New A Rev.}%
\def\pasa{PASA}%
\def\pra{Phys.~Rev.~A}%
\def\prb{Phys.~Rev.~B}%
\def\prc{Phys.~Rev.~C}%
\def\prd{Phys.~Rev.~D}%
\def\pre{Phys.~Rev.~E}%
\def\prl{Phys.~Rev.~Lett.}%
\def\pasp{PASP}%
\def\pasj{PASJ}%
\def\qjras{QJRAS}%
\def\rmxaa{Rev. Mexicana Astron. Astrofis.}%
\def\skytel{S\&T}%
\def\solphys{Sol.~Phys.}%
\def\sovast{Soviet~Ast.}%
\def\ssr{Space~Sci.~Rev.}%
\def\zap{ZAp}%
\def\nat{Nature}%
\def\iaucirc{IAU~Circ.}%
\def\aplett{Astrophys.~Lett.}%
\def\apspr{Astrophys.~Space~Phys.~Res.}%
\def\bain{Bull.~Astron.~Inst.~Netherlands}%
\def\fcp{Fund.~Cosmic~Phys.}%
\def\gca{Geochim.~Cosmochim.~Acta}%
\def\grl{Geophys.~Res.~Lett.}%
\def\jcp{J.~Chem.~Phys.}%
\def\jgr{J.~Geophys.~Res.}%
\def\jqsrt{J.~Quant.~Spec.~Radiat.~Transf.}%
\def\memsai{Mem.~Soc.~Astron.~Italiana}%
\def\nphysa{Nucl.~Phys.~A}%
\def\physrep{Phys.~Rep.}%
\def\physscr{Phys.~Scr}%
\def\planss{Planet.~Space~Sci.}%
\def\procspie{Proc.~SPIE}%
\let\astap=\aap
\let\apjlett=\apjl
\let\apjsupp=\apjs
\let\applopt=\ao





\normalem
\begin{acknowledgements}
 We would like to acknowledge the Aryabhatta Research Institute of observational sciencES (ARIES), Nainital for making  telescope time  available. We would also like to acknowledge the Herschel Science Archive from which we have downloaded \emph{Herschel} SPIRE 500$\mu m$ map of CB27. We collected \emph{SCUBA} 850$\mu m$ map from the CADC repository of the SCUBA Polarimeter Legacy catalogue and is greatly acknowledged. This work has made use of data from the European Space Agency (ESA) mission {\it Gaia} (\url{https://www.cosmos.esa.int/gaia}), processed by the {\it Gaia}
Data Processing and Analysis Consortium (DPAC,
\url{https://www.cosmos.esa.int/web/gaia/dpac/consortium}). Funding for the DPAC has been provided by national institutions, in particular the institutions participating in the {\it Gaia} Multilateral Agreement. The anonymous reviewer of this paper is highly acknowledged for his/her comments and suggestions which definitely helped to improve the quality of the paper.
The author G. B. Choudhury acknowledges the funding agency Department of Science and Technology (DST), Government of India for providing DST INSPIRE fellowship (IF 170830).

\end{acknowledgements}


\bibliographystyle{raa}
\bibliography{bibtex}

\end{document}